\documentstyle[aps,12pt,bbm,epsfig,aps,prl,twocolumn,doublespace,amsfonts]{revtex}
\begin{document}
%
%
%
%
\onecolumn
\title{Exotic Statistics for Ordinary Particles in Quantum Gravity }
\author{John Swain\\ March 31, 2008}
\address{Department of Physics, Northeastern University, Boston, MA 02115, USA\\
email: john.swain@cern.ch}
\date{Awarded an Honorable Mention in the Gravity Research Foundation Essay Competition, 
2008}
\maketitle

\begin{abstract}
\section*{\bf Summary}
 Objects exhibiting statistics other
than the familiar Bose and Fermi ones are natural in theories
with topologically nontrivial objects including geons, strings,
and black holes. It is argued here from several viewpoints
that the statistics of {\em ordinary} particles with which
we are already familiar are likely to be modified due to
quantum gravity effects. In particular, such modifications
are argued to be present in loop quantum gravity and in any
theory which represents spacetime in a fundamentally piecewise-linear
fashion. The appearance of unusual statistics
may be a {\em generic} feature (such as the deformed
position-momentum uncertainty relations and the appearance
of a fundamental length scale) which are to
be expected in any theory of quantum gravity, and which
could be testable.
\end{abstract}

\clearpage

\section{Introduction}

The spin-statistics theorem\cite{SandW} states that 
half-integer spin particle obey Fermi-Dirac statistics
and the Pauli exclusion principle and integer spin
particles obey Bose-Einstein statistics is well-known
and has many derivations\cite{Paulibook}. I argue in this
essay that lifting the restrictions in the usual derivations
of this theorem leads one to expect {\em generically} that
it will not hold when quantum gravitational effects are
included. In fact, I will give so many arguments for this
that I argue that this may well be a {\em generic} feature
of quantum theories of gravity\cite{generic}. Conversely,
if quantum gravity effects leave the usual field commutators
and spin-statistics connection untouched, then something very deep
must be going on to protect them.

\section{The Standard Approaches to the Spin-Statistics Theorem}

There are two main ways to approach the spin-statics
theorem. The first is to consider commutation relations for
fields $\phi(t,{\bf x})$ and their conjugate momenta
$\pi(t,{\bf x})$ and {\em assume} that it makes sense to think of
them as being analogous to $x$ and $p$ of nonrelativistic
quantum mechanics. Suppressing Lorentz indices, one has
expressions of the form

\begin{eqnarray}
\left[\phi(t,{\bf x}),\phi(t,{\bf x}^\prime)\right] = 0 \\
\left[\pi(t,{\bf x}),\pi(t,{\bf x}^\prime)\right] = 0 \\
\left[\phi(t,{\bf x}),\pi(t,{\bf x}^\prime)\right] = i\delta^{(3)}({\bf x} - {\bf x}^\prime) 
\end{eqnarray}
where the brackets can represent commutators or
anticommutators. Ignoring interactions and assuming locality
and flat Minkowski spacetime, one then finds that one
runs into problems if commutators are used for Dirac (half integer
spin) fields or if anticommutators are used for Bose
(integer spin) fields.

\begin{figure}[htbp]
\begin{center}
\hspace*{-5mm}\mbox{\epsfig{file=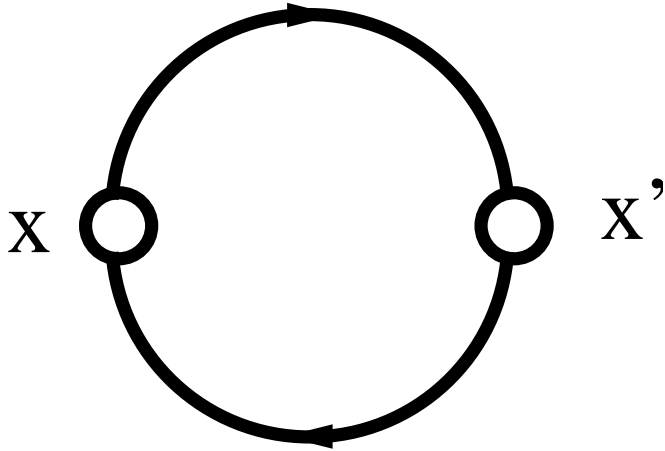,height=6.0cm}}
\end{center}
\caption{Two particles (small circles) swapped by a rotation
(two arcs with arrows). Here no flux passes through the circle
that defines the rotation, but if it did, one might expect
additional phases to appear.}
\label{fig:exchange}
\end{figure}

An alternative viewpoint makes the swap of
${\bf x}$ and ${\bf x^\prime}$ more physical by arguing 
\cite{Berry} that it is equivalent to a rotation (FIG. \ref{fig:exchange}).

If one thinks of two particles
being swapped as equivalent to a rotation of $2\pi$, then
one picks up a sign of $e^{(i2\pi s)}$ for spin
$s$: a ``$+$'' for bosons and a ``$-$'' for fermions.
Implicit in such an argument is that there
are no gauge fields present which could have altered the phases
of particles {\it en route}.

\section{Effects of Quantum Gravity}

The most obvious concern with equations (1), (2), and (3) above is
that they are equal-time commutation relations with
an obvious reliance on a 3+1 split -- something which might be
dealt with by considering Peierls brackets instead \cite{Peierls}.
Of course one would still have to worry about
possible extensions (central or not) when replacing brackets
with commutators in the quantum theory.

Rather general physical arguments
suggest that the commutators between $x$ and $p$
should be modified by quantum gravitational effects \cite{GUP},
as should the $x$ commutators\cite{Ahluwalia1}.
\footnote{I would argue for nontrivial commutation relation for positions
already without quantum gravity.\cite{me-NCG}}. The momentum
commutators are already nontrivial in curved spacetime. 

For fields,
the $\delta^{(3)}({\bf x}-{\bf x}^\prime)$ of (3) might be replaced by
a sharply-peaked function with width related to the Planck mass. It is
worth noting that such scale-dependent effects arise naturally for
composite particles. The idea that statistics of composite particles
might be subtle goes back to Wigner\cite{Wigner} in 1929 and
Ehrenfest and Oppenheimer\cite{Ehrenfest} in 1931. For fermion
pairs (``quasibosons'' \cite{quasibosons}) such as superfluid helium-4
and Cooper pairs a short-range Pauli effect {\em is} present.
Lipkin\cite{Lipkin} notes that the energy gap for Cooper pairs ``...would be
absent if the fermions behaved like simple bosons.'' In other words,
we already know of systems which behave more or less bosonic or fermionic
as a function of scale.

The most convincing general arguments for 
changes to the basic commutators are perhaps those of 
Ahluwalia-Khalilova\cite{ahlu} based on the work of
Mendes, Chryssomalakos and Okon
\cite{MCO}. The idea is to find the most general stable
extension of the 
combined Poincar\'{e} and Heisenberg algebras. This leads to 
the Snyder-Yang-Mendes algebra\cite{SYM}, which 
has nontrivial commutators (not all commutation relations
are shown)
between coordinates $X_\mu$ and momenta $P_\mu$:

\begin{eqnarray}
\left[X_\mu,X_\nu\right] = i\ell_P^2 J_{\mu\nu} \\
\left[P_\mu,P_\nu\right] = i\frac{\hbar^2}{\ell_C^2} J_{\mu\nu} \\ 
\left[P_\mu,X_\nu\right] = i\hbar\eta_{\mu\nu}F + i\hbar\beta J_{\mu\nu}
\end{eqnarray}
These involve $\hbar$ (explicit here),
two length scales $\ell_P$ (presumably of order
the Planck length) and $\ell_C$ (presumably of cosmological size),
a new dimensionless constant $\beta$, and the angular momentum $J_{\mu\nu}$.
$F$ is a new operator having nontrivial commutation
relations with $P$ and $X$. The triply
special relativity of Kowalski-Glikman
and Smolin\cite{TSR} can be put in this form\cite{ahlu}.

Given that one {\em postulates} field commutators by analogy with the
commutation relations of $x$ and $p$, it is now by no means
obvious that (1),(2) and (3) are correct.

One can also stick to the usual commutation relations for the
Poincar\'e and Heisenberg group and ask questions at the level 
of the fields themselves. After all, physically (and in the spirit
of noncommutative geometry\cite{NCG}), one {\em constructs} spacetime
from measurements involving fields.

Now consider the product $\phi(x)\phi(y)$ which is needed to 
form commutators, anticommutators and propagators ($x$ and $y$ now
commuting labels for spacetime points).
The first problem is that $\phi(x)\phi(y)$ is not gauge invariant.
As noted long ago by Schwinger, $\phi(x)\phi(y)$ would need to be
multiplied by a phase $\exp(i\int_x^y A_\mu dx^\mu)$ for whatever
connection $A$ is relevant.
For a self-interacting charged particle
in flat spacetime one finds\cite{myfrac} that the free infrared propagator
$1/(p^2-m^2)$ is raised to a fractional power ($1+\frac{\alpha}{\pi}$) and
becomes nonlocal (a charged particle carries a long-range field). 
It has been argued that Newtonian gravitational self-interaction even in
flat spacetime will give a similar sort of correction\cite{myfrac}.

In a general curved space background\cite{Birrel-and-Davies},
but ignoring self-interaction,
DeWitt \cite{DeWitt75BD} has given an exact representation
for the Feynman propagator which is complicated and nonlocal
\footnote{Even then there
are subtleties.Toms has argued\cite{Toms} that there is an ambiguity due to the choice
of path integral measure.}. Of course all these discussions have
assumed adiabatic processes where particle numbers do not change 
as in the Unruh and Hawking effects \cite{Birrel-and-Davies}.

Since what originally looked like a 
well-defined product (or expectation value thereof) of two
free fields in flat spacetime
must be replaced for interacting fields in curved
spacetime by a nonlocal object, it seems
unlikely that their (anti-)commutators would suffer no modifications.

The foregoing two arguments, one kinematical and the other
dynamical, suggest that the usual commutation relations
for quantum fields (and thus the usual derivations of the spin-statistics
connection) may not be generally valid. 
There have been arguments in the literature 
that there should be no unusual spin-statistics relationships
in curved spacetime\cite{nogo}.
They require assumptions about the (anti-)commutation relations, 
as well as the existence 
of suitable flat regions of spacetime. Bardek {\em et al.}\cite{Bardek}
considered exotic statistics in curved spacetime and argued for
their consistency and constancy in an expanding universe, while Scipioni
has argued\cite{Scipioni}
that transitions of statistics might
occur under some conditions. A comprehensive list
of references can be found in\cite{NIST}.

At first glance, statistics describes
exchanges of objects (at the particle level) and thus is 
about representations of the permutation group of $n$ objects $S_n$.
If an exchange is
equivalent to a rotation \cite{Berry}, and one wants two exchanges
to be the identity, then one is naturally led to the representation
of a single exchange by
multiplying a wavefunction by $\pm 1$. This was formalized
with path-integral arguments by Laidlaw and
deWitt\cite{LdeW} who ruled out any other possibilities -- but there
are loopholes.

Extended objects are well-known to support
exotic statistics. Topological geons\cite{Freedman-Sorkin}
can violate the spin-statistics theorem\cite{geonstats}.
Charged extremal black holes have been argued\cite{Strominger}
to obey ``infinite statistics''\cite{DHR}, where any representation of the permutation 
group can occur.
Infinite statistics admit a Fock-like representation\cite{Greenberg-et-al} 
in terms of deformed commutation
relations for creation and annihilation operators of
the form $a_ka_l^\dag-qa_l^\dag a_k=\delta_{kl}$. A theory of objects
satisfying these deformed commutation relations has a Hamiltonian
which is nonlocal and nonpolynomial in the field operators -- something
obviously suggestive of curved spacetime and the expected nonlocalities
described above.

If the particles have internal degrees of freedom ({\it i.e.} their wavefunctions
are sections of ${\mathbb{C}}^N$, or something else) there can be
inequivalent quantizations, labelled by irreducible representations
of the braid group $B_n(M)$. All sorts of novel statistics are then
possible \cite{Imbo1,Imbo2}. Intuitively, internal degrees of freedom can keep
track of how many times particles have gone ``around each other''.

In 2+1 dimensions, it is well-known that particles coupled
to gauge fields can form composite objects (``anyons'') with unusual
statistics\cite{exoticstats}.
The physical idea is easy to understand. If one takes a particle
around a flux $\Phi$, one picks up an Aharonov-Bohm type
phase $\exp(ig\Phi)$ where $g$ represents the coupling of the
particle to the field. The braid group $B_n$ keeps track
of how the flux lines twist around each other as the particles
move in 2 dimensions. Gravitational anyons in 2+1 dimensions\cite{Grav-anyons}
could be physically relevant at the surfaces of black holes (perhaps
even microscopic or virtual ones\cite{regulator}).

It has been suggested \cite{fieldang} that
exotic statistics for charged particles ({\em i.e.} some
electrons in a white dwarf acting as bosons) could arise for
particles whose angular momentum comes partly from coupling
to external electromagnetic fields. Similar effects might
be anticipated from gravitational fields.

Strings open up completely new possibilities. For a space $M$ like ``${\mathbb{R}}^3$
with n points removed'' (say by black holes or some sort of spacetime foam) one
could have $\pi_1(M)=0$ but strings would probe the loop space $\Omega M$, with
$\pi_1(\Omega M)=\pi_2(M)\neq 0$. 

Exotic statistics are also possible\cite{stringstats}  for strings even in
topologically trivial 3-dimensional manifolds, due to their ability to
be linked and tangled. For this, a further generalization of braid
statistics is needed involving the ``loop braid group'' LB$_n$\cite{Baez}.
Lest one imagine that such issues are purely academic, Niemi has argued for
exotic statistics in ``leapfrogging'' vortex rings\cite{Niemi}
in quantum liquids and gases which, like anyons, {\em really exist} in the physical
world.  Exotic statistics for strings in 4-D BF theory have also been discussed by Baez, Wise,
and Crans\cite{Baez}. Particles corresponding to states of strings might  
inherit exotic statistics which could test stringy models
of particles. That said, stringiness can also be somewhat hidden in the nonlocality of
theories of point particles coupled to long-range fields \cite{particlesasstrings}.

Gambini and Setaro\cite{GS} found fractional statistics for composites
of charged one-dimensional objects and vortices. Fort and Gambini found Fermi-Bose
transmutation \cite{FG1} for point scalars and Nielsen-Olesen strings in the
Maxwell-Higgs system, and fractional statistics\cite{FG2} in a 3+1 dimensional system composed
of an open magnetic vortex and an electrical point charge.
 
\section{Exotic Statistics for Ordinary Particles in LQG}

I now want to concentrate on a very physical reason
why loop quantum gravity (LQG)\cite{LQG}
and any 3+1 formulation of 
a piecewise-linear (PL)\cite{discrete} Regge-like\cite{Regge,DT} theory
of gravity naturally supports exotic
statistics. 

The idea is very simple: 
in 3+1 dimensional Regge-type theories\cite{3dregge}, the curvature is
distributional, with support on {\em edges} where flat 3-simplices meet.
A particle picks up a phases as it moves around a {\em line}
of singular curvature and that phase can contribute to exotic statistics.

\begin{figure}[htbp]
\begin{center}
\hspace*{-5mm}\mbox{\epsfig{file=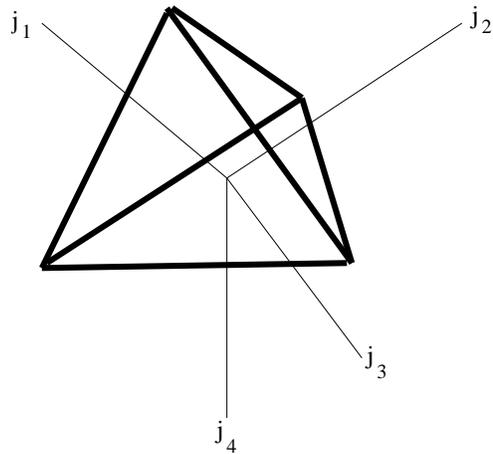,height=6.0cm}}
\end{center}
\caption{A portion of a spin network and an associated simplex.
The areas of the faces of the simplex are determined by the SU(2) 
representations j$_i$ on the edges which pierce them and the
volume by the intertwiner associated with the vertex.}
\label{fig:simplex}
\end{figure}

\begin{figure}[htbp]
\begin{center}
\hspace*{-5mm}\mbox{\epsfig{file=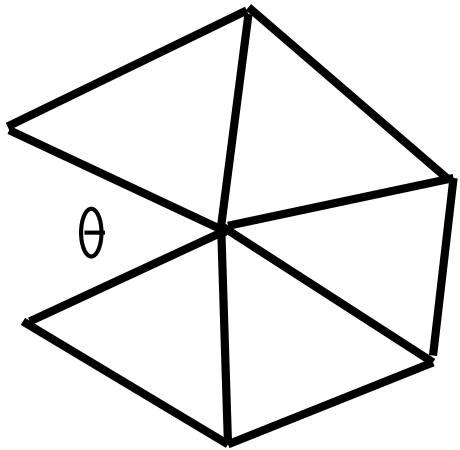,height=6.0cm}}
\end{center}
\caption{View of 2-simplices fitting together with a deficit
angle (the edges there should be brought together to construct the
curved 2-geometry with curvature at the central point). 
The 3-dimensional picture is analogous, but
the 5 triangles are now replaced by 5 tetrahedra from a spin-network
and the point where the curvature is becomes a line of curvature.
One could think of this figure as looking ``down'' on the 5 flat
triangular faces of 5 flat tetrahedra joined along a line of
singular curvature perpendicular to the page.}
\label{fig:regge1}
\end{figure}

In LQG, spin-networks define natural
dual PL simplicial geometries \cite{otherPL,me-lqg} of flat simplicial complexes
with distributional curvature along 1-dimensional subspaces where
they join.
A spin network is a graph composed
of lines carrying $SU(2)$ representation labels which give areas to
surfaces they pierce. They meet at vertices labelled by intertwiners.
If there are at least 4 edges which meet at a vertex, one can think
of the vertex as enclosed by faces which get their areas from the edges
that pierce them. The intertwiner determines the volume enclosed (see 
FIG. \ref{fig:simplex} for an example). 
Curvature is represented by the deficit angle along edges where
simplices meet (see FIG. \ref{fig:regge1} for an example).

If the distributional lines of curvature are taken to represent matter,
we have physical strings and all the arguments for the loop braid group
also appear in this context.

While I cannot claim to know the correct replacement for the usual
(anti-)commutation relations in field theory,
I would argue that their modification is a logical
and {\em generic} possibility in theories of quantum gravity and
worthy of theoretical and experimental study. It is a natural extension
of the now familiar ideas of looking for changes to the basic
commutators of $x$ and $p$. Such modifications might only appear at
very large or very small scales ($\ell_C$ or $\ell_P$) and could
easily have avoided experimental limits so far\cite{Mohapatra}.

I would like to conclude with some speculations connecting ideas
discussed here with other topics of current research. If one thinks
of local supersymmetry transformations as local changes of statistics
of objects which can be fermions or bosons, then local changes of
symmetry lead naturally to local translation invariance and
thus to general coordinate invariance. Is there a connection here?
Jackson\cite{Jackson} has argued that a suitable
position-momentum commutator can describe many features of gravity. 
To make this plausible, recall that essentially all the
interesting things about the geometry of phase space in
quantum mechanics come from the one nontrivial commutator.

Braiding and exotic statistics may
also have some bearing on interpretations of Standard
Model particles in terms of framed spin networks\cite{BT,BTothers}.
Framings arise naturally in spin-networks with q-deformed
groups\cite{qdefLQG} which are needed in LQG with a
cosmological constant (nontrivial $\lbrack P_\mu, P_\nu\rbrack$).
Also of interest is\cite{Anandan}
in which spin and statistics for spacetime and ``internal''
exchanges are connected. It is also interesting that non-commutative
geometries arise naturally together with q-deformed groups in 
the same situations where anyons appear\cite{Floratos}, so it
seems many ideas may be connected.

\section{Acknowledgements}

I would like to thank Ka\'{c}a Bradonji\'{c} and Tom
Paul for careful readings of drafts of this paper. I
would also like to thank all Dharam
Ahluwalia for having brought the issue of stability of a deformed
Poincar\'e-Heisenberg algebra to my attention, and
various members of the loop quantum gravity community for having
said interesting things at one point or another at LOOPS '07,
especially Lee Smolin, Fotini Markopoulou, Seth Major and Sundance
Bilson-Thompson. This work was supported in part by the US National
Science Foundation.

{\em Note added:} The day after this work was completed, a preprint
from Mark G. Jackson
entitled ``Spin-Statistics Violations from Heterotic String Worldsheet
Instantons'' appeared on Arxiv (ArXiv:0803.4472v1) which discusses
possible violations of the spin-statistics theorem in heterotic
string theory and also makes the point that should the true scale
for quantum gravity be much lower than the usual Planck scale, 
such effects might be more readily observable in the near future.


\begin{thebibliography}{99}
\bibitem{SandW} R. F. Streater and A. S. Wightman, ``PCT, Spin and
Statistics, and All That'', Princeton University Press, 2000.
\bibitem{Paulibook} I. Duck and E. C. G. Sudarshan, ``Pauli and the
Spin-Statistics Theorem'', World Scientific, March 1998.
\bibitem{generic} L. Smolin, ``Generic Predictions of Quantum Theories of Gravity'', hep-th/0605052
\bibitem{Berry} M. V. Berry and J. M. Robbins, Proc. Roy. Soc. {\bf A453} (1997) 1771.
\bibitem{GUP} There is a large literature on generalizing the uncertainty principle,
which in a string theory concept goes back at least to Veneziano: G. Veneziano,
Europhys. Lett. {\bf 2} (3) (1986) 199. For a review, see F. Scardigli, Phys. Lett. {\bf B452}
(1999) 39.
\bibitem{Ahluwalia1} D. V. Ahluwalia, Physics Letters {\bf B339} (1994) 301
(Honorable Mention in the 1994 Awards for Essays in Gravitation)
\bibitem{Peierls} See for example, G. Bimonte, G. Esposito, G. Marmo, and C. Stornaiolo,
Int. J. Mod. Phys. {\bf A18} (2003) 2033 and references therein.
\bibitem{me-NCG} S. Sivasubramanian, G. Castellani, N. Fabiano, A. Widom, J. Swain,
Y.N. Srivastava, G. Vitiello, 
Annals Phys. {\bf 311} (2004) 191-203;
S. Sivasubramanian, G. Castellani, N. Fabiano,
A. Widom, J. Swain, Y. N. Srivastava, and G. Vitiello,
J. Mod. Optics, {\bf 51} (2004) 1529.
\bibitem{Wigner} E. P. Wigner, Math. und Naturwiss. Anzeiger der Ungar. Ak. der Wiss.
{\bf 46} (1929) 576 (cited in \cite{quasibosons})
\bibitem{Ehrenfest} P. Ehrenfest and J. R. Oppenheimer, Phys. Rev. {\bf 37} (1931) 333.
\bibitem{quasibosons}  W. A. Perkins, Int. J. Theor. Phys. {\bf 41} (2002) 823. 
\bibitem{Lipkin} H. J. Lipkin, ``Quantum Mechanics'', North-Holland, Amsterdam, 1973, Ch. 6
(quoted in \cite{quasibosons}).
\bibitem{ahlu} D. V. Ahluwalia-Khalilova, Class. Q. Grav. {\bf 22} (2005) 1433.
\bibitem{MCO} R. V. Mendes, J. Phys. {\bf A27} (8091);
C. Chryssomalakos and E. Okon, Int. J. Mod. Phys. {\bf D13} (2004) 2003.
\bibitem{SYM} H. S. Snyder, Phys. Rev. {\bf 71} (1947) 38;
C. N. Yang, Phys. Rev. {\bf 72} (1974) 874;
R. V. Mendes, J. Phys. {\bf A27} (1994) 8091;
R. V. Mendes, J. Math. Phys. {\bf 41} (2000) 156;
\bibitem{TSR} J. Kowalski-Glikman and L. Smolin, Phys. Rev.{\bf D70} (2004) 065020.
\bibitem{NCG} A. Connes, ``Noncommutative Geometry'', Academic Press, 1994. 
\bibitem{myfrac} S. Gulzari, J. Swain, and A. Widom,
Mod. Phys. Lett. {\bf 21} (2006) 2861-2871;
S. Gulzari, Y. N. Srivastava, J. Swain, and A. Widom, Proceedings of IRQCD, June 5--9, 2006,
Rio de Janeiro, and Braz. J. Phys. vol. 37, no. 1b. March 2007, page 286.
\bibitem{Birrel-and-Davies} N. D. Birrell and P. C. W. Davies,``Quantum fields in curved space'',
Cambridge University Press, 1982.
\bibitem{DeWitt75BD} B. S. De Witt, ``The Dynamical Theory of Groups and Fields'',
in {\em Relativity, Groups and Topology}, eds. B. S. DeWitt and C. DeWitt,
Gordon and Breach, 1965; Phys. Rep. {\bf 19C} (1975) 297.
\bibitem{Toms} D. J. Toms, ``The Schwinger Action Principle and the Feynman Path
Integral for Quantum Mechanics in Curved Space'', hep-th/0411233.
\bibitem{nogo} L. Parker and Y. Wang, Phys. Rev. {\bf D39} (1989);
J. W. Goodson and D. J. Toms, Phys. Rev. Lett. {\bf 71} (1993) 3240;
R. Verch, Commun. Math. Phys. {\bf 223} (2001) 261.
\bibitem{Bardek} V. Bardek, S. Meljanac, and A. Perica, Phys. Lett. {\bf B338} (1994) 20.
\bibitem{Scipioni} R. Scipioni, Il Nuov. Cim., {\bf 112B} (1997) 119.
\bibitem{NIST}{ \tt http://physics.nist.gov/MajResFac/EBIT/peprefs.html}
\bibitem{LdeW} M. G. G. Laidlaw and C. Morette DeWitt, Phys. Rev. {\bf D3} (1971) 1375.
\bibitem{Freedman-Sorkin} J. L. Friedman and R.Sorkin, Phys. Rev. Le
{\bf 44} (1980) 1100; {\bf 45} (1980) 148, General Relativity and
Gravitation {\bf 14} (1982) 615.
\bibitem{geonstats} A. P. Balachandran, E. Batista, I. P. Costa e
Silva and P. Teotonio-Sobrinho, Nuclear Physics {\bf B566} (2000) 441;\\
C. Anezeris, A. P. Balachandran, M. Bourdeau, S. Jo, T. R. Ramadas,
and R. D. Sorkin, Mod. Phys. Lett. {\bf A4} (1989) 331;
C. Aneziris, A. P. Balachandran, M. Bourdeau, S. Jo, R. D. Sorkin, and T. R. Ramadas
Int. J. Mod. Phys. {\bf A4} (1989) 5459.
\bibitem{Strominger} A. Strominger, Phys. Rev. Lett., {\bf 71} (1993) 3397.
\bibitem{DHR} S. Doplicher, R. Haag, and J. Roberts, Comm. Math. Phys.
{\bf 23} (1971) 199: {\bf 35} (1974) 49
\bibitem{Greenberg-et-al} O. W. Greenberg, Phys. Rev. Lett. {\bf 64} (1990)
705; Phys. Rev. {\bf D43} (1991) 4111.
\bibitem{Imbo1} T. D. Imbo and E. C. G. Sudarshan, Phys. Rev. Lett. {\bf 60} (1988) 481.
\bibitem{Imbo2} T. D. Imbo, C. S. Imbo, E. C. G. Sudarshan, Phys. Lett. {\bf B234} (1990) 103.
\bibitem{exoticstats} F. Wilczek, Phys. Rev. Lett. {\bf 48} (1982) 1144;
Phys. Rev. Lett. {\bf 49} (1982) 957; R, Mackenzie and F. Wilczek,
Int. J. Mod. Phys. A3 (1988) 2827.
\bibitem{Grav-anyons} S. Deser, Phys. Rev. Lett. {\bf 1990}, Y. M. Cho, D. H. Park, and C. G. Han, 
Phys. Rev. {\bf D43} (1991) 1421.
\bibitem{regulator} L. Crane and L. Smolin, Gen. Rel. Grav., {\bf 17} (1985) 1209.
\bibitem{qdefLQG} S. Major and L Smolin, Nucl. Phys. {\bf B473} (1996) 267.
\bibitem{fieldang} A. Kato, G. Mu\~{n}oz, D. Singleton,  J. Dryzek, and V. Dzhunushaliev,
Found. Phys. {\bf 33} (2003) 769; S. Mandal and S.Chakrabarty, ``Electrons as quasi-bosons
in Strong Magnetic Fields and the Stability of Magnetars'', astro-ph/0209462;
J. Dryzek, A. Kato, D. Mu\~{n}oz, and D. Singleton, ``Electrons as quasi-bosons in magnetic
white dwarfs'', astro-ph/0110320, D. Singleton and J. Dryzek, ``Electromagnetic field
angular momentum in condensed matter systems'', cond-mat/0009068.
\bibitem{stringstats} J. A. Harvey and J. Liu, Phys. Lett. {\bf B240} (1990) 369;
X. Fustero, R. Gambini, and A. Trias, Phys. Rev. Lett. {\bf 62} (1989) 1964;
C. Aneziris, A. P. Balachandran, L. Kauffman, and A. M. Srivastava,
Int. J. Mod. Phys. {\bf A6} (1991) 2519, 
C. Aneziris, Mod. Phys. Lett. {\bf A7} (1992) 3789;
S. Surya, J. Math. Phys. {\bf 45} (2004) 2515.
\bibitem{Baez} J. C. Baez, D. K. Wise, and A. S. Crans, ``Exotic Statistics for Strings
in 4d BF theory'', gr-qc/0603085, 9 May 2006, published
in Advances in Theoretical and Mathematical Physics, vol. 11., No. 5, October 2007
\bibitem{Niemi} A. J. Niemi, Phys. Rev. Lett. {\bf 94} (2005) 124502
\bibitem{particlesasstrings} P.-M. Ho, Phys.Lett. {\bf B558} (2003) 238.
\bibitem{GS} R. Gambini and L. Setaro, Phys. Rev. Lett. {\bf 65} (1990) 2623.
\bibitem{FG1} H. Fort and R. Gambini, Phys. Lett. {\bf B372} (1996) 226.
\bibitem{FG2} H. Fort and R. Gambini, Phys. Rev. {\bf D54} (1996) 1778.
\bibitem{LQG} See for example, C. Rovelli,``Quantum Gravity'',
Cambridge University Press, 2004 and
T. Thiemann, ``Modern Canonical Quantum General Relativity'',
Cambridge University Press, 2007.
\bibitem{discrete} T. Regge and R. M. Williams, J. Math. Phys.
{\bf 41} (2000) 3964.
\bibitem{Regge} T. Regge, Nuov. Cim. {\bf 19} (1961) 558.
\bibitem{DT} J. Ambj{\o}rn, M. Carfora, and A. Marzuoli,
``The Geometry of Dynamical Triangulations'', Lecture Notes in 
Physics, Springer, Berlin, 1997. 
\bibitem{3dregge}T. Piran and R. M. Williams, Phys. Rev. {\bf D33}
(1986) 1622.
\bibitem{me-lqg} J.Swain, talk at Loops '07 and paper in preparation.
\bibitem{otherPL} F. Markopoulou, gr-qc/9704013.
\bibitem{Mohapatra} O. W. Greenberg and R. N. Mohapatra, Phys. Rev. {\bf D39} (1989) 2032. 
\bibitem{Jackson} M. G. Jackson, Int. J. Mod. Phys. {\bf D14} (2005) 2239
(Honorable Mention in the 2005 Awards for Essays in Gravitation)
\bibitem{BT} S. O. Bilson-Thompson, hep-ph/0503213
\bibitem{BTothers} S. O. Bilson-Thompson, F. Markopoulou and L. Smolin, Class.Quant.Grav. {\bf 24} (2007)
3975-3994.
\bibitem{Anandan} J. Anandan, Phys. Lett. {\bf A248} (1998) 124.
\bibitem{Floratos} E. G. Floratos, ``Bohm-Aharonov Interactions and q-quantum mechanics'',
Proceedings of the Lepton-Photon Symposium, 1991, vol. 1, 107.
\end{thebibliography}
\end{document}